\documentstyle[osa,manuscript,psfig]{revtex}

\begin{document}

\titlepage
\title{
\hskip 11.0truecm {{\normalsize FUB-HEP/95-14}}\\
\vspace {1.2truecm}
Hyperon production mechanisms and \\
single-spin asymmetry in high energy
hadron-hadron collisions
}
\author {C. Boros and Liang Zuo-tang}
\address {Institut f\"ur theoretische Physik,
Freie Universit\"at Berlin \\
Arnimallee 14, 14195 Berlin, Germany}
\maketitle

\begin{abstract}                

It is shown that the existence of left-right asymmetry in
single-spin inclusive Lambda production,
together with the characteristic features of the data,
should be considered as another clear signature for the existence
of orbiting valence quarks in polarized nucleons.
Predictions for other hyperons are made.
It is pointed out that
measurements of such asymmetries are very helpful
not only for probing the spin structure
of the nucleons but also for studying
the production mechanisms of the hyperons.

\end{abstract}

\newpage

Striking left-right asymmetries $(A_N)$
have been observed
in single-spin inclusive production processes
at high energy (200 GeV/c)
now not only for pions[1]
but also for $\Lambda $[2] and $\eta $[3].
Comparing the new data[2,3] with
those for pions[1] we see that,
while the data[3] for $\eta $ is
essentially the same as that for $\pi ^0$,
that for $\Lambda $ as a function of
$x_F(\equiv 2p_\|/\sqrt{s}$,
where $p_\| $ is the longitudinal
momentum of the produced hadron
with respect to the
incident direction of the projectile, and
$s$ is the total center of
mass energy of the colliding system squared.)
shows the following:
(1) Similar to those for pions,
$A_N(x_F,\Lambda|s )$ is significant
in the large $x_F$ region ($x_F\agt 0.6$, say),
but small in the central region.
(2) $A_N(x_F,\Lambda |s )$ is negative
in the large $x_F$ region, and
it behaves similarly as
$-A_N(x_F,\pi ^+|s)$ does in this region.
(3) $A_N(x_F,\Lambda |s)$ begins to rise up
later than $A_N(x_F,\pi^\pm |s)$ does.
For $x_F$ in the neighborhood
of $x_F\sim 0.4-0.5$,
it is even positive.
In the region $0<x_F\alt 0.5$,
its behavior is similar to
that of $A_N(x_F,\pi^0 |s)$.

In recent publications[4,5],
we pointed out that
the existence of left-right
asymmetry for pion production
should be considered as a strong
indication for the
existence of orbiting
valence quarks in polarized
nucleons.
Left-right asymmetries are expected to exist
in inclusive single-spin processes when
direct formations of particles (or particle systems)
through fusions and/or annihilations
of the orbiting valence quarks of
the polarized projectile with suitable antiquarks
associated with the target contribute significantly.
Single-spin inclusive experiments are extremely
useful in probing the orbital motion of the valence
quarks in polarized hadrons.
They provide also useful information
for studying the production mechanisms of mesons
in general and for distinguishing those mesons
from direct formation processes
from the others in particular.
In connection with the
new data[2,3] from E704 Collaboration,
we note the following:
While that for $\eta $ was expected in our earlier
publications[4,5], no discussion has yet been made on
hyperon production.
We are therefore led to the following questions:
Does direct formation process play
a significant role
also for Lambda- and/or other hyperon-production
in the fragmentation regions?
Can we also understand the left-right asymmetry
for Lambda production, especially
the above mentioned characteristics of the data?
What do we expect to see for other hyperons?

These are the questions we would like to discuss in this note.
We show that the first two questions
should be answered in the affirmative and that
predictions for other hyperons can be made.
In fact, as we will see in the following,
just as mesons,
for moderately large transverse momentum
($p_\perp \agt 0.6$ GeV/c, say),
a significant part of hyperons
observed in the fragmentation regions are products of
direct formation processes.
The above-mentioned differences and similarities
between $A_N(x_F,\Lambda |s)$
and $A_N(x_F,\pi |s)$ directly reflect the display
between the different kinds of direct formation processes
and the non-direct-formation part.
The existence of $A_N(x_F,\Lambda |s)$,
together with its above-mentioned characteristics,
should be considered as another clear signature
for the existence of orbiting valence quarks in polarized nucleon.
Measurements of such left-right asymmetries
are helpful not only in probing the
spin structure of the nucleon but also in
studying the production mechanisms of the hyperons.

We begin our discussions by
briefly summarizing the key points of
the proposed picture,
which have been applied successfully to
meson and lepton-pair productions in [4,5],
as follows:

(i) Valence quarks in a nucleon
are treated as Dirac particles
in an effective confining potential.
It is shown that orbital motion
for these valence quarks
is always involved even when they are in
the ground states,
and that the direction of the orbital motion is determined
uniquely by the polarization of the valence quark.

(ii) Valence quarks in a polarized hadron
are polarized either in the
same or in the opposite direction as the hadron.
This is determined by the wave function of the hadron.
For proton, on the average,
5/3 of the $u$ valence quarks
are polarized in the same and 1/3 of
them in the opposite direction as the proton.
For $d$, they are 1/3 and 2/3 respectively.

(iii) A significant part of the particles
(with moderately large transverse momenta)
observed in the fragmentation regions
in high-energy hadron-hadron
collisions are products of the direct formation
processes of the valence quarks of
one of the colliding hadrons with suitable
anti-sea-quarks associated with the other.

(iv) There exists a significant surface effect
in single-spin inclusive production processes.
This implies that only those particles directly formed near
the front surface of the polarized hadron
retain the information of the polarization.
This leads to the conclusion that, for example,
a meson which is formed through direct fusion
of an upward polarized valence quark with an anti-sea-quark
should get an extra transverse momentum from the orbital motion of the
valence quark to the left (looking down stream, as defined in the
experiments[1,2,3]).

We note that points (i), (ii) and (iv)
are independent of
what kind of particles we look at.
They are valid also when we study the
single-spin inclusive hyperon production processes.
We recall that (iii) is a well-known fact
for lepton-pair production, and is consistent with the
data analysis by Ochs[6],
the theoretical calculations by Das and Hwa[7],
and the recent calculations from us[8] for meson production.
How is the situation for hyperon production?
We recall that in proton we have two $u$
and one $d$ valence quarks,
and that the flavor content of Lambda is $uds$.
Hence, there are following three possibilities
for direct formations for $\Lambda$-production:

(a) A $(u_vd_v)$-valence-diquark
from the projectile $P$ picks up
a $s_s$-sea-quark associated with the target
$T$ and forms a $\Lambda $:
$(u_vd_v)^P+s_s^T\to \Lambda $.

(b) A $u_v$-valence-quark
from the projectile $P$ picks up a
$(d_ss_s)$-sea-diquark associated
with the target $T$ and forms a $\Lambda $:
$u_v^P+(d_ss_s)^T\to \Lambda $.

(c) A $d_v$-valence-quark
from the projectile $P$ picks up a
$(u_ss_s)$-sea-diquark associated with
the target $T$ and forms a $\Lambda $:
$d_v^P+(u_ss_s)^T\to \Lambda $.

We first consider their contributions to
Lambda production in unpolarized reaction
$p(0)+p(0)\to \Lambda +X$
and compare them with data[9].
[Here, as well as in the following of this paper, we use
the same notations as we used in [4,5]:
The first particle on the left-hand-side
of a reaction denotes the projectile,
the second is the target,
$(0)$ means unpolarized,
and $(\uparrow )$ means transversely polarized.]
Just as that for meson[4,5],
the number density for the $\Lambda $ produced through
these three direct formation processes
can be expressed as,
\begin{equation}
D_a(x_F,\Lambda |s)=\kappa_{\Lambda}^d
f_D(x^P|u_vd_v) s_s(x^T),
\end{equation}
\begin{equation}
D_b(x_F,\Lambda |s)= \kappa_{\Lambda}
   u_v(x^P) f_D(x^T|d_ss_s),
\end{equation}
\begin{equation}
D_c(x_F,\Lambda |s)= \kappa_{\Lambda}
d_v(x^P) f_D(x^T|u_ss_s),
\end{equation}
respectively.
Here $x^P\approx x_F$ and $x^T\approx m_\Lambda ^2/(sx_F)$,
followed from
energy-momentum conservation in the direct formation processes.
$q_i(x)$ is the quark distribution function, where $q$ denotes
the flavor of the quark and the subscript $i$ denotes whether
it is for valence or sea quarks.
$f_D(x|q_iq_j)$ is the diquark distribution functions,
where $q_iq_j$ denotes the flavor and whether they are
valence or sea quarks.
$\kappa _\Lambda ^d$ and $\kappa _\Lambda $ are two constants.
The number density for Lambda produced in an unpolarized
reaction $p(0)+p(0)\to \Lambda +X$ is given by:
\begin{equation}
N(x_F,\Lambda |s)=N_0(x_F,\Lambda |s) +D(x_F,\Lambda |s).
\end{equation}
where $D(x_F,\Lambda |s)=\sum_{i=a,b,c}D_i(x_F,\Lambda |s)$ is
the total contribution from direct formation processes.
Using the parametrizations of
the quark and diquark distribution
given in the literatures[10,11],
we calculated $D(x_F,\Lambda |s)$.
The two constants $\kappa _\Lambda $ and $\kappa _\Lambda ^d$
are fixed by fitting two data points
in the large $x_F$-region.
The results are compared with the data[9] in Fig.~1.
It is interesting to see that the data
can indeed be fitted very well in the fragmentation region.
In fact, the characteristic feature of the data[9],
compared with those for pions[12],
is that it is much broader than the latter,
and this is just a direct consequence of the contribution
from the valence diquarks through
the process (a) given above.
We see also that
the whole $0<x_F<1$ region can be divided into three parts:
In the large $x_F$-region (say, $x_F\agt 0.6$),
the direct process (a) plays the dominating role;
and for small $x_F$-values ($x_F\alt 0.3$, say),
the non-direct-formation part $N_0(x_F,\Lambda|s)$
dominates, while in the middle
(that is, in the neighborhood of $x_F\sim 0.4-0.5$),
the direct formation processes (b) and (c)
provide the largest contributions.

Having seen that the unpolarized
data[9] in the fragmentation region
can indeed be described using the
direct formation mechanism,
we continue to discuss the left-right
asymmetry in the corresponding processes
with transversely polarized proton projectile.
We note that according to points (i), (ii) and (iv),
$\Lambda $ produced through
the direct formation process (b)
should have large probabilities to go left
and thus give positive contributions to $A_N$,
while those from $(c)$ contribute negatively to it.
For the direct formation process (a),
we note the following:
This direct formation process (a) should be
predominately associated with the production of
a meson directly formed through fusion of
the $u$ valence quark of the projectile
with a suitable anti-sea-quark of the target.
It follows from points (i),(ii) and (iv)
(see also [4,5]) that this meson should
have a large probability to
obtain an extra transverse momentum to the left.
Thus, according to momentum conservation,
the Lambda produced through (a) should
have a large probability to
obtain an extra transverse momentum to the right.
This implies that $(a)$ contributes negatively to $A_N$,
opposite to that of the associatively produced meson
($\pi ^+$ or $K^+$ or other).
As we have mentioned above,
(a) plays the dominating role
in the large $x_F$ ($x_F\agt 0.6$) region.
We therefore expect that
$A_N(x_F,\Lambda |s)$
is large in magnitude and negative in sign,
and behaves in a similar way as
$-A_N(x_F,\pi^+ |s)$ does in this region.
In the region $x_F\alt 0.5$,
we have the display between the
non-direct-formation part $N_0(x_F,\Lambda |s)$ and
the contributions from
the direct formation processes (b) and (c).
The situation is very much similar to
that for $\pi ^0$-production.
We expect therefore that, for $x_F\alt 0.5$,
$A_N(x_F,\Lambda |s)$ is positive in sign, smaller than
$A_N(x_F,\pi^+ |s)$, and has the similar behavior as
$A_N(x_F,\pi^0 |s)$.
It should changes its sign somewhere and rises up
later than $-A_N(x_F,\pi^+ |s)$ does.
All these qualitative features are consistent
with the characteristics of the data[2]
from the E704 Collaboration.

Encouraged by these good agreements between the qualitative
results and the characteristics of the data, we now
calculate $A_N(x_F,\Lambda |s)$ quantitatively.
We recall that left-right asymmetry is defined as,
\begin{equation}
A_N(x_F,\Lambda|s)\equiv \frac{N(x_F,\Lambda|s,\uparrow)-
N(x_F,\Lambda|s,\downarrow)}
{N(x_F,\Lambda|s,\uparrow)+N(x_F,\Lambda|s,\downarrow)} ,
\end{equation}
where
\begin{equation}
N(x_F,\Lambda |s,i)\equiv \frac{1}{\sigma_{in}}
 \int_R d^2p_{\perp} \frac{d\sigma}{dx_F dp^2_{\perp} }
 (x_F,\vec p_{\perp},\Lambda|s,i),
\end{equation}
(with $i=\uparrow $ or $\downarrow $)
is the normalized number density of
the  observed $\Lambda$ in a given
kinematical region $R$ when
the projectile is upwards ($\uparrow$)
or downwards ($\downarrow$) polarized.
$\sigma_{in}$ is the total
inelastic cross section.
$x_F\equiv 2 p_\|/\sqrt{s}$,
$p_\|$ and $p_\perp $ are
the longitudinal and transverse
components of the momentum of the Lambda.
According to the proposed picture,
$\Delta N(x_F,\Lambda |s)\equiv
N(x_F,\Lambda |s,\uparrow\nobreak)
- N(x_F,\Lambda|s,\downarrow)$
comes predominately from the direct formation part
of the $\Lambda $'s and is proportional to
$\Delta D(x_F,\Lambda |s)\equiv D(x_F,\Lambda |s,+)-D(x_F,\Lambda |s,-)$
for the case (b) and (c).
Here, $D(x_F,\Lambda |s,\pm)$ denotes the number density of the
$\Lambda $'s formed through direct fusion or combination of
valence quarks [$u_v$ in case (b) and $d_v$ in case (c)]
polarized in the same $(+)$ or opposite $(-)$
direction as the projectile with
suitable sea-diquarks associated with the target.
The contribution from (a) is opposite in sign to that
of the associatively produced meson and is proportional
to $- r(x|u_v,tr)\equiv -\Delta u_v(x|tr)/u_v(x)$,
where $x$ is the fractional momentum of the $u_v$ valence quark.
We have therefore,
\begin{equation}
\Delta N(x_F,\Lambda |s)=C \Bigl [
-r(x|u_v,tr) D_a(x_F,\Lambda|s) +
\Delta D_b(x_F,\Lambda|s) +
\Delta D_c(x_F,\Lambda|s) \Bigr ].
\end{equation}
The expressions for
$D_{b,c}(x_F,\Lambda |s,\pm )$ are
similar to those for
$D_{b,c}(x_F,\Lambda |s)$ which are
given by Eqs. (2) and (3) with $u_v(x^P)$
and $d_v(x^P)$ being replaced by
$u_v^\pm(x^P|tr)$ and $d_v^\pm(x^P|tr)$.
Here,  $u_v^\pm(x^P|tr)$ and $d_v^\pm(x^P|tr)$ are the
number densities of the valence quarks polarized in the
same $(+)$ or opposite $(-)$ direction as the
transversely polarized proton.
We obtain therefore,
{\small
\begin{equation}
A_N(x_F,\Lambda |s)=C
{-\kappa _\Lambda ^d r(x|u_v,tr) f_D(x^P|u_vd_v)s_s(x^T)+
\kappa _\Lambda  [\Delta u_v(x^P|tr) f_D(x^T|d_ss_s)+
\Delta d_v(x^P|tr) f_D(x^T|u_ss_s)]
\over N(x_F,\Lambda |s) },
\end{equation}
}
where $\Delta q_v(x|tr)\equiv q_v^+(x|tr)-q_v^-(x|tr)$.
For a rough estimation of $A_N(x_F,\Lambda|s)$,
we use the ansatz for
$q_v^\pm (x|tr)$ used in [4,5],
which is the simplest one
that satisfies
the condition mentioned in point (ii).
In this case $r(x|u_v,tr)=2/3$ is a constant.
The results of this calculation
are compared with the data[2] in Fig.~2.
We see that all the qualitative
features of the data[2] are well reproduced.

Similar analysis can be made for
other hyperons in a
straight-forward way.
For example, for $\Sigma ^+$,
which has a flavor content $uus$,
we have following two
possibilities for direct formation:
$(u_vu_v)^P+s_s^T\to \Sigma ^+$, and
$u_v^P+(u_ss_s)^T\to \Sigma ^+$.
The first process contributes to $A_N$
opposite to that from the associatively produced
meson through fusion of $d_v$ with
a suitable anti-sea-quark, and is thus
opposite to $A_N(x_F,\pi^-|s)$
or similar to $A_N(x_F,\pi^+|s)$.
The contribution from the second one
is determined by the $u_v$
valence quark and is also similar
to $A_N(x_F,\pi^+|s)$. We therefore expect that
$A_N(x_F,\Sigma ^+|s)$ has the similar
behavior as $A_N(x_F,\pi^+|s)$
in the whole $x_F$ region $0<x_F<1$.
For $\Sigma ^-$, whose flavor content is $dds$,
we have only one possibility for
direction formation, i.e. $d_v^P+(ds)_s^T\to \Sigma ^-$.
We expect therefore $A_N(x_F,\Sigma ^-|s)$ to
have the same behavior as
$A_N(x_F,\pi^-|s)$.
Although $\Sigma^0$ has the same flavor content as $\Lambda$,
there is the following difference:
While the $ud$-valence-diquark in $\Lambda $ is in
the state with the sum of
their total angular momenta $j_{ud}=0$,
the $ud$-valence-diquark
in $\Sigma ^0$ is in a $j_{ud}=1$ state.
We note that the proton wave function[5]
can be written in the following way[14],
\begin{equation}
|p(\uparrow) \rangle ={1\over 2\sqrt{3}} \Bigl \{
3u(\uparrow)\cdot {1\over \sqrt{2}}
\Bigl [ u(\uparrow) d(\downarrow)-u(\downarrow) d(\uparrow) \Bigr ]
+ u(\uparrow)\cdot {1\over \sqrt{2}}
\Bigl [ u(\uparrow) d(\downarrow)+u(\downarrow) d(\uparrow) \Bigr ]
-\sqrt {2} u(\downarrow) u(\uparrow) d(\uparrow) \Bigr \},
\end{equation}
where $(\uparrow )$ and $(\downarrow )$ denote that the
$z$-component of the total angular momentum of the corresponding
valence quark is either $j_z=+1/2$ or $j_z=-1/2$ respectively.
We see clearly that, if the $ud$-diquark is in a
$j_{ud}=1$-state, the other $u$-valence-quark is either
polarized upward or downward,
with relative probabilities $1:2$.
So if $\Sigma^0 $ is produced through the same kind
of direct formation process as shown in (a),
the associatively produced meson should have a large
probability to go right.
Hence, we expect that $A_N(x_F,\Sigma^0|s)$ behaves differently
from $A_N(x_F,\Lambda|s)$ does
in the large $x_F$ region ($x_F\agt 0.6$, say).
In contrast to $A_N(x_F,\Lambda|s)$,
$A_N(x_F,\Sigma ^0|s)$ is positive
in sign in this region.
But, they have the similar behavior in the $x_F\alt 0.5$ region.
This implies also that there is no change of sign in
$A_N(x_F,\Sigma ^0|s)$ as a function of  $x_F$.
Similarly, we expect that
$A_N(x_F,\Xi ^0|s)$ behaves similar to $A_N(x_F,\pi ^+|s)$;
while $A_N(x_F,\Xi ^-|s)$ likes $A_N(x_F,\pi ^-|s)$.
In table I, we summarize the sign of these
$A_N$'s in the large $x_F$ region for different
hyperons.

It should be emphasized that,
although the left-right asymmetries
are in general different for different hyperons,
they have the following in common:
All of them are significant mainly in the
fragmentation region of the polarized colliding object,
and are approximately independent of the unpolarized one.
No asymmetry is expected for hyperons in the fragmentation
region of the unpolarized colliding object.
All these can be tested by future experiments.
It is also interesting to see that
the polarization of hyperons observed[15]
in reactions with
unpolarized projectile and unpolarized target
are also significant mainly
for sufficiently large $p_\perp $ and/or
in the fragmentation region.
It can be imagined that
such hyperon polarizations
and the left-right asymmetries
in single-spin processes are closely related to each other.
Such a study is under way, the results will be published
elsewhere[16].

In summary, we have shown
that the striking left-right asymmetry for
inclusive Lambda production in single-spin
hadron-hadron collisions[4]
is a direct consequence of
the existence of orbiting valence quarks in polarized nucleons.
Direct formation of particles through fusions of valence quarks
(diquarks) with suitable sea-anti-quarks (sea-quarks) plays a
significant role not only for the
production of mesons but also for hyperons
with moderately large $p_\perp $
in the fragmentation region.
This implies that left-right asymmetry exists
not only for $\Lambda$ but also for other hyperons
in single-spin inclusive production processes.
Measurements of such asymmetries provide extremely useful
information not only for the spin structure of nucleon
but also for the production mechanisms of hyperons.

We thank Meng Ta-chung and
R. Rittel for helpful discussions.
This work was supported in part by Deutsche
Forschungsgemeinschaft (DFG:Me 470/7-1).

\newpage

\begin{center}
\begin{table}[h]
\caption { Sign for the left-right asymmetry
$A_N$ for hyperon production in
$p(\uparrow )+p(0)\to $Hyperon + X in the
large $x_F$ region. }
\begin{tabular}{|l|c|c|c|c|c|c|}
\hline
&\phantom {tttt} &\phantom {tttt}
&\phantom {tttt}  &\phantom {tttt}
&\phantom {tttt}  &\phantom {tttt} \\[-0.2truecm]
Hyperons\ \  & $\Lambda $ &
$\Sigma^+$ & $\Sigma^0$ &$\Sigma^-$ & $\Xi^0$ & $\Xi^-$ \\[0.2truecm]
\hline
& & & & & & \\[-0.2truecm]
 $A_N(x_F\agt 0.6)$ & $-$ & $+$ & $+$ & $-$ & $+$ & $-$ \\[0.2truecm]
\hline
\end{tabular}
\end{table}
\end{center}

\newpage

\noindent
{\large Figures}
\vskip 0.6truecm
\noindent
Fig.1:
The differential cross section $E d\sigma /dp^3$ for
$p(0)+p(0)\to \Lambda + X$
as a function of $x_F$
at $p_\perp = 0.65$ GeV/c and ISR-energies
as a sum of different contributions.
The dash-dotted and the two dotted lines
represent the contributions from the
direct formation processes (a), (b) and
(c) given in the text respectively.
The dashed line represents the
non-direct-formation part, which is parametrized as
$300(1-x_F)^2e^{-3x_F^3}$.
The solid line is the sum of all contributions.
The data is taken from Ref. [9].

\vskip 0.6truecm
\noindent
Fig.2:
Left-right asymmetry $A_N$ for $p(\uparrow)+p(0)\to
\Lambda +X$ at $200$ GeV/c.
The data are from Ref. [2] and the low energy data
are from Ref. [13].


\begin{thebibliography}{mmm}
\bibitem{[14]} It was not necessary to differentiate the
$j_{ud}=0$ and $j_{ud}=1$ states when calculating
$A_N(x_F,\Lambda|s)$. This is because,
although it is true that $ud$-diquark in a
$j_{ud}=1$ state combines with a $s$-quark and
forms a $\Sigma ^0$, $\Sigma ^0$ decays
into $\Lambda $ (and a photon). Since the mass difference
between $\Lambda $ and $\Sigma^0 $ is relatively small,
the momentum and the corresponding left-right asymmetry
of $\Sigma ^0$ will be mainly transferred to $\Lambda $.
The results remain approximately the same,
as if the $\Lambda$ were directly formed.
\end{thebibliography}
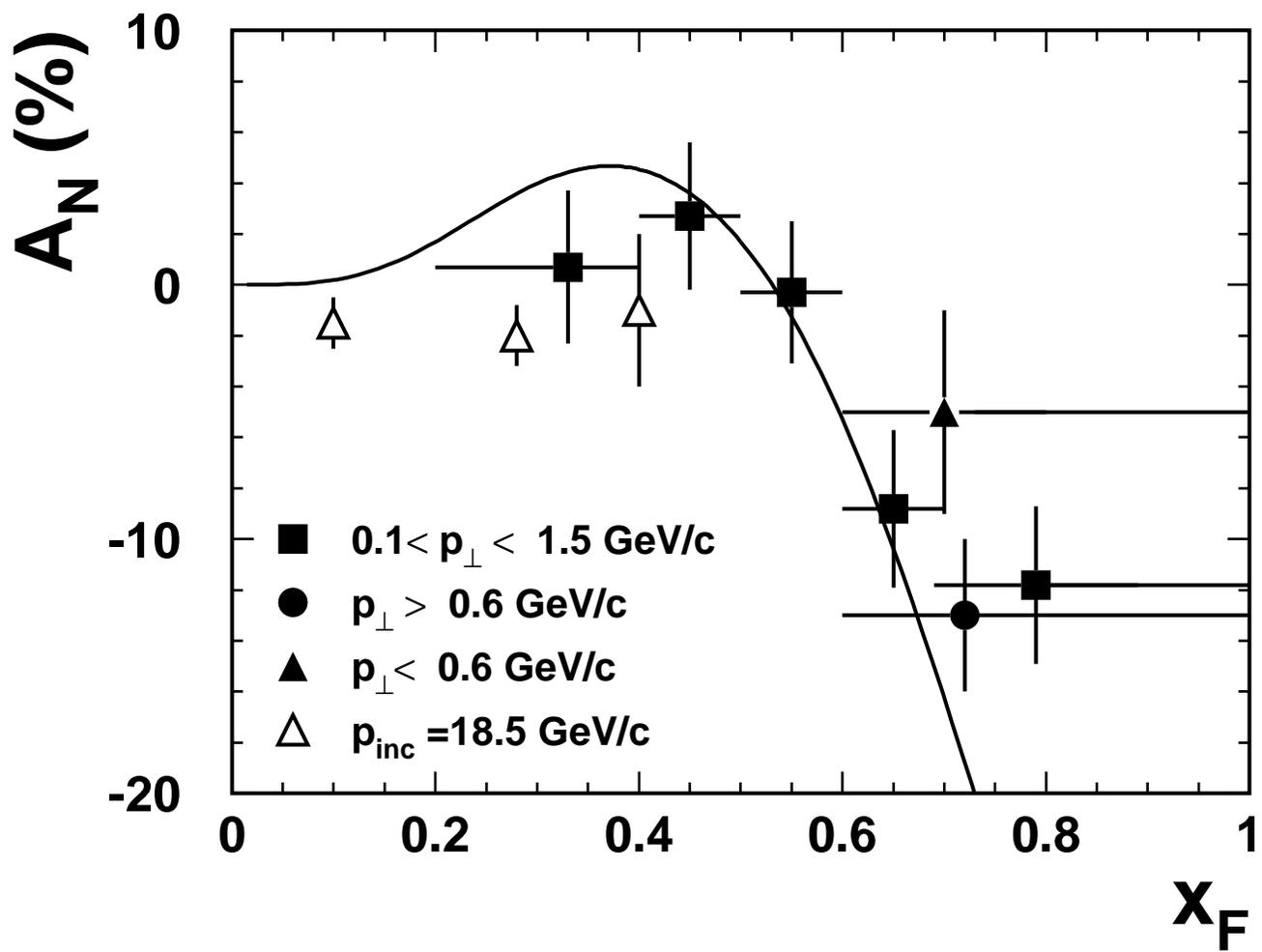
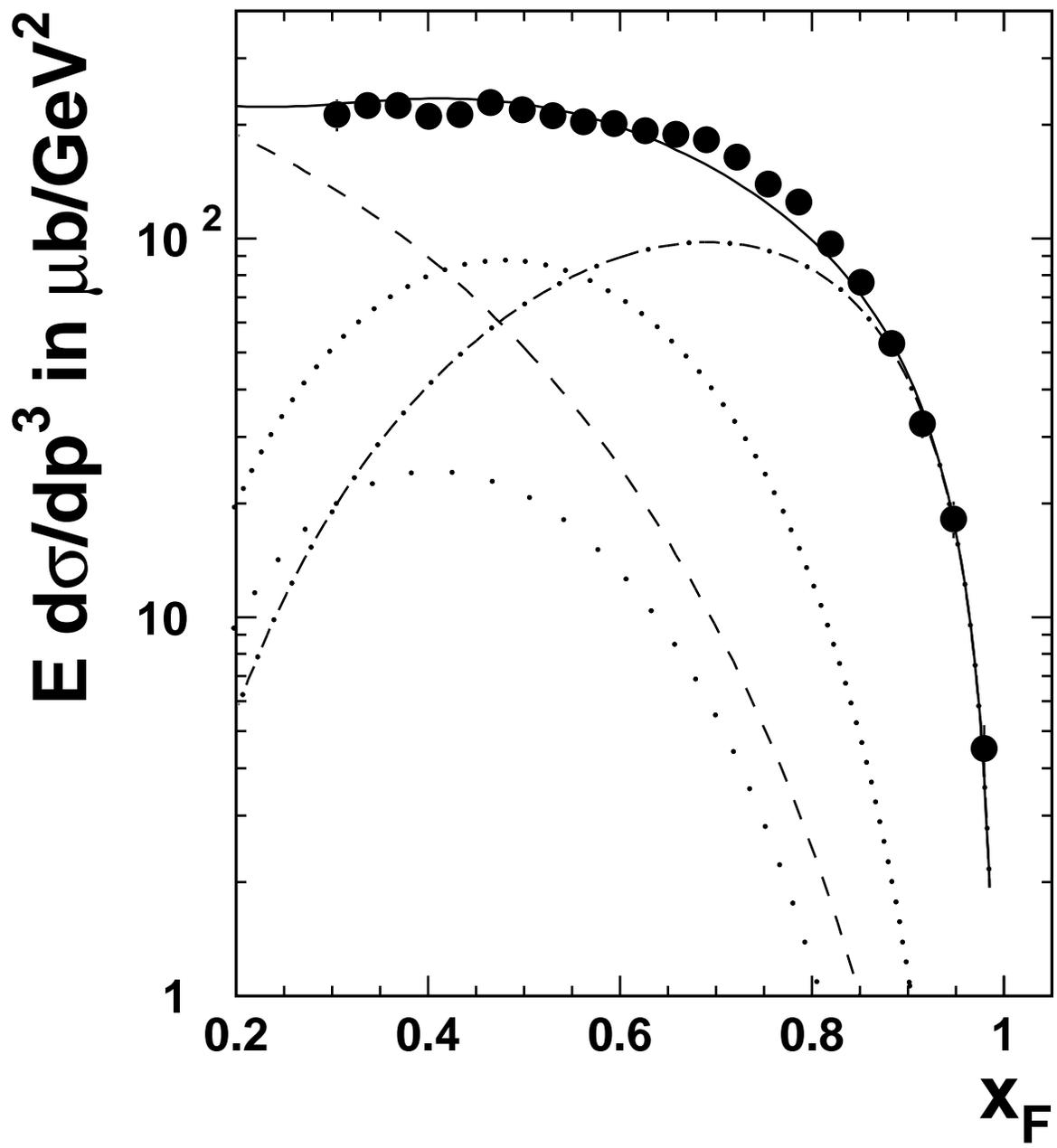
\end{document}